\documentclass[aps,prb,amsmath,amssymb, superscriptaddress, preprint]{revtex4-1}
\usepackage[utf8]{inputenc} 

\usepackage{geometry} 
\usepackage{graphicx} 
\usepackage{color}

\usepackage{booktabs} 
\usepackage{array} 
\usepackage{paralist} 
\usepackage{verbatim} 
\usepackage{subfig} 
\usepackage{hyperref}

\newcommand*{\citen}[1]{%
  \begingroup
    \romannumeral-`\x 
    \setcitestyle{numbers}%
    \cite{#1}%
  \endgroup   
}

\makeatletter
\newcommand*{\rom}[1]{\expandafter\@slowromancap\romannumeral #1@}
\makeatother

\begin{document}

\newcommand{\beq}{\begin{equation}}
\newcommand{\bequo}{\begin{quotation}}
\newcommand{\beqa}{\begin{eqnarray}}
\newcommand{\eeq}{\end{equation}}
\newcommand{\equo}{\end{quotation}}
\newcommand{\eeqa}{\end{eqnarray}}
\newcommand{\non}{\nonumber}
\newcommand{\mx}{\mbox}
\newcommand{\mxf}[1]{\mbox{\footnotesize{#1}}}
\newcommand{\lb}{\label}
\newcommand{\fr}[1]{(\ref{#1})}
\newtheorem{entry}{}[section]
\newcommand{\bent}[1]{\vspace*{-1.5cm}\hspace*{-1cm}\begin{entry}\lb{e{#1}}\rm}
\newcommand{\eent}{\end{entry}}
\newtheorem{discussion}{Discussion}[section]
\newcommand{\bdisc}[1]{\vspace*{-1.cm}\begin{discussion}\lb{d{#1}}\rm }
\newcommand{\edisc}{\end{discussion}}
\newcommand{\hs}{\hspace*{5mm}}

\newcommand{\fig}[1]{\lb{f{#1}}}
\newcommand{\frf}[1]{\ref{f{#1}}}

\newcommand{\frr}[1]{\href{run:./BlueRefs/#1}}

\newcommand{\fre}[1]{{\bf\ref{e{#1}}}}
\newcommand{\frd}[1]{{\bf\ref{d{#1}}}}
\newcommand{\Emark}{$\sqcap\hspace{-2.7mm}\sqcup$}
\newcommand{\fn}{\footnote}

\newcommand{\capt}{\caption}

\renewcommand{\a}{\alpha}
\renewcommand{\b}{\beta}
\newcommand{\g}{\gamma}
\newcommand{\G}{\Gamma}
\renewcommand{\d}{\delta}
\newcommand{\s}{\sigma}
\renewcommand{\S}{\Sigma}
\renewcommand{\th}{\theta}
\newcommand{\Th}{\Theta}
\newcommand{\D}{\Delta}
\newcommand{\e}{\epsilon}
\newcommand{\w}{\omega}
\newcommand{\W}{\Omega}
\newcommand{\al}{\alpha}
\newcommand{\bet}{\beta}
\newcommand{\gam}{\gamma}
\newcommand{\lam}{\lambda}
\newcommand{\Lam}{\Lambda}
\newcommand{\eps}{\epsilon}

\newcommand{\red}{\color{red}}
\newcommand{\blue}{\color{blue}}
\newcommand{\green}{\color{green}}
\definecolor{gray}{rgb}{0.5, 0.5, 0.5}
\newcommand{\gray}{\color{gray}}
\definecolor{ro}{rgb}{1, .5,0.}
\newcommand{\ro}{\color{ro}}
\definecolor{ora}{rgb}{1, .75,0.}
\newcommand{\ora}{\color{ora}}
\definecolor{oy}{rgb}{1, .85,0.4}
\newcommand{\oy}{\color{oy}}
\definecolor{yel}{rgb}{.95, .9,.55}
\newcommand{\yel}{\color{yel}}
\definecolor{yg}{rgb}{0.5, .9,.4}
\newcommand{\yg}{\color{yg}}
\definecolor{gr}{rgb}{0.35, .75,.45}
\newcommand{\gr}{\color{gr}}
\definecolor{bg}{rgb}{0.20, 0.6,0.75}
\newcommand{\bg}{\color{bg}}
\definecolor{bl}{rgb}{0.2, 0.6,1}
\newcommand{\bl}{\color{bl}}
\definecolor{bp}{rgb}{0.6, 0.3,1}
\newcommand{\bp}{\color{bp}}
\definecolor{pur}{rgb}{0.7, 0.,.9}
\newcommand{\pur}{\color{pur}}
\definecolor{rp}{rgb}{0.9, 0.,.7}
\newcommand{\rp}{\color{rp}}

\definecolor{str}{rgb}{1, .3,.6}
\newcommand{\str}{\color{str}}
\definecolor{brown}{rgb}{.7, .4,0.}
\newcommand{\brown}{\color{brown}}

\newcommand{\see}{$\rightarrow$}
\newcommand{\react}{$\longrightarrow$}
\newcommand{\equil}{$\longleftrightarrow$}
\newcommand{\map}{\rightarrow}
\newcommand{\pam}{\leftarrow}
\newcommand{\maps}{\rightarrow}
\newcommand{\imply}{\Rightarrow}
\newcommand{\Lrar}{\Leftrightarrow}
\newcommand{\rar}{\rightarrow}
\newcommand{\irrev}{\prec\prec}
\newcommand{\adb}{\stackrel{\mbox{\fns A}}{\sim}}
\newcommand{\go}{\prec}
\newcommand{\adeq}{\stackrel{A}\sim}
\newcommand{\theq}{\stackrel{T}\sim}

\newcommand{\pder}[2]{\frac{\partial {#1}}{\partial {#2}}}
\newcommand{\pdert}[2]{\frac{\partial^2 {#1}}{\partial {#2}^2}}
\newcommand{\fder}[2]{\frac{\delta {#1}}{\delta {#2}}}
\newcommand{\PDD}[3]{\left.\frac{\partial^{2}{#1}}{\partial{#2}^{2}}\right|_{#3}}
\newcommand{\PD}[3]{\left(\frac{\partial{#1}}{\partial{#2}}\right)_{#3}}
\newcommand{\der}[2]{\frac{d {#1}}{d {#2}}}
\newcommand{\Tr}{\mbox{Tr}\,}

\newcommand{\upu}{$^{+}$}
\newcommand{\upd}{$^{2+}$}
\newcommand{\upt}{$^{3+}$}
\newcommand{\upq}{$^{4+}$}
\newcommand{\umu}{$^{-}$}
\newcommand{\umd}{$^{2-}$}
\newcommand{\umt}{$^{3-}$}
\newcommand{\umq}{$^{4-}$}
\newcommand{\sz}{$_0$}
\newcommand{\su}{$_1$}
\newcommand{\sd}{$_2$}
\newcommand{\st}{$_3$}
\newcommand{\sq}{$_4$}
\renewcommand{\sc}{$_5$}
\newcommand{\sse}{$_6$}
\newcommand{\ssi}{$_7$}
\newcommand{\so}{$_8$}
\newcommand{\pa}{$\a$}
\newcommand{\pb}{$\b$}
\newcommand{\pg}{$\g$}
\newcommand{\pd}{$\d$}
\newcommand{\pe}{$\epsilon$}
\newcommand{\ps}{$\sigma$}

\newcommand{\AH}{\mbox{H}}
\newcommand{\AC}{\mbox{C}}
\newcommand{\AO}{\mbox{O}}
\newcommand{\AN}{\mbox{N}}
\newcommand{\AP}{\mbox{P}}
\newcommand{\ANa}{\mbox{Na}}
\newcommand{\AK}{\mbox{K}}
\newcommand{\ACa}{\mbox{Ca}}
\newcommand{\AB}{\mbox{B}}
\newcommand{\AFe}{\mbox{Fe}}
\newcommand{\ACu}{\mbox{Cu}}
\newcommand{\AAl}{\mbox{Al}}
\newcommand{\ASi}{\mbox{Si}}
\newcommand{\AMg}{\mbox{Mg}}
\newcommand{\ASe}{\mbox{Se}}
\newcommand{\AAs}{\mbox{As}}
\newcommand{\AS}{\mbox{S}}
\newcommand{\AHe}{\mbox{He}}
\newcommand{\AAr}{\mbox{Ar}}
\newcommand{\AXe}{\mbox{Xe}}
\newcommand{\ACl}{\mbox{Cl}}
\newcommand{\AF}{\mbox{F}}
\newcommand{\ABr}{\mbox{Br}}
\newcommand{\AI}{\mbox{I}}

\title{Numerical study of non-adiabatic quantum thermodynamics of the driven resonant level model: Non-equilibrium entropy production and higher order corrections}

\author{Kaiyi Tong}
\affiliation{School of Science, Westlake University, Hangzhou, Zhejiang 310024, China}
\affiliation{Institute of Natural Sciences, Westlake Institute for Advanced Study, Hangzhou, Zhejiang 310024, China}

\author{Wenjie Dou}
\email{douwenjie@westlake.edu.cn}
\affiliation{School of Science, Westlake University, Hangzhou, Zhejiang 310024, China}
\affiliation{Institute of Natural Sciences, Westlake Institute for Advanced Study, Hangzhou, Zhejiang 310024, China}

\begin{abstract}

We present our numerical study on quantum thermodynamics of the resonant level model subjected to non-equilibrium condition as well as external driving. Following our previous work on non-equilibrium quantum thermodynamics (Phys. Rev. B 101, 184304 [2020]), we expand the density operator into a series of power in the driving speed, where we can determine the non-adiabatic thermodynamic quantities. Particularly, we calculate the non-equilibrium entropy production rate as well as higher order non-adiabatic corrections to the energy and/or population. In the limit of weak system-bath coupling, our results reduce to the one from the quantum master equation. 



\end{abstract}

\maketitle

\section{Introduction} \label{sec:intro}

The study of dynamics and thermodynamics for a quantum system strongly coupled to a set of baths are of great interests recently,\cite{gemmer2009quantum,kosloff2013quantum, PhysRevLett.114.080602,vinjanampathy2016quantum,anders2017focus,PhysRevLett.116.240403,alicki2018introduction,benenti2017fundamental,RevModPhys.83.771,PhysRevLett.102.210401,liu2021periodically} particularly due to its applications in nano technology, quantum information, and quantum measurement.\cite{poot2012mechanical, pekola2015towards, rossnagel2016single,PhysRevE.96.052106,PhysRevLett.122.110601} Quantum thermodynamics address the energy and information flow of a system consisting of a few atoms or qubits interacting strongly to non-equilibrium environments.\cite{mishaPRBfriction,millen2016perspective, gemmer2009quantum, talkner2007fluctuation,brandao2015second,jarzynski2011equalities} Moreover, to make a useful quantum engine, external driving is usually applied. The finite speed driving can introduce  non-adiabatic effects, including entropy production, friction (or dissipation), and fluctuation. \cite{solinas2013work,schmidt2015work,gogolin2016equilibration,subacsi2012equilibrium,ness2017nonequilibrium,PhysRevX.7.021003}

The driven resonant-level model has been studied extensively for thermodynamics in the strongly coupled regimes.\cite{mishaPRBfriction,PhysRevLett.114.080602,PhysRevB.93.115318,haughian2018quantum} At equilibrium (with one bath or no current), studies based on different methods (e.g symmetric splitting\cite{PhysRevB.93.115318,PhysRevB.94.035420}, scattering matrix\cite{bruch2018landauer}, and non-equilibrium Green's function\cite{PhysRevLett.114.080602}) arrive at  a (somewhat) consistent quantum description of the thermodynamics. In particular, the first order non-adiabatic corrections to work, population, and entropy have been identified. Furthermore, the entropy production is proportional  to the frictional work and remains positive at equilibrium, which is consistent with the second law of thermodynamics. That being said, out of equilibrium (baths with different temperatures or chemical potentials), no such formulations have been agreed upon.\cite{semenov2020transport,bergmann2021green,strasberg2021first,bruch2018landauer} Particularly the definition and the positivity of entropy production remain as open questions. Furthermore, there are far less studies on the higher order non-adiabatic corrections to the thermodynamic quantities, which is a challenging task.

In the previous studies,\cite{PhysRevB.98.134306,dou2020universal} by expanding the density operator into a series of power in the driving speed, we formulate a general description of quantum thermodynamics for a generic model strongly coupled to one or more baths. We identify the non-adiabatic corrections to thermodynamic quantities, such as work, population, and entropy. In particular, the non-adiabatic entropy production rate is given by\cite{dou2020universal} 
\begin{eqnarray}
\delta \dot S = -k_B \sum_{\alpha\nu}  \int_0^{\infty} \dot R_{\alpha}  \dot R_{\nu} tr(  e^{-i \hat H t'/\hbar} \partial_\nu \hat  \rho_{ss} e^{i \hat  H t'/\hbar} \partial_\alpha \ln \hat \rho_{ss} ) dt' .
\end{eqnarray}
Here, $\hat \rho_{ss}$ is the steady state density of the total system. $\hat H$ is the total Hamiltonian, which depends on a set of external parameters $\{R_\alpha\}$.  $\dot R_\alpha$ is the driving speed. The above equation can be recasted into a Kubo transformed correlation function using the Barker-Campell-Hassdrauff formula. Such that we have proven that the entropy production rate is always positive, which is consistent with the second law of thermodynamics. At equilibrium, the steady state density reduces to equilibrium density, such that we can show that the non-adiabatic entropy is proportional  to the friction (or non-adiabatic correction to the work). Out of equilibrium, however, we have not been able to calculate $\delta \dot S$ explicitly due to the difficulty of determining $\hat \rho_{ss}$. 



In the present manuscript, we calculate this entropy production rate explicitly for the driven resonant level model, where we discretize the continuous bath degrees of freedom and diagonalize the total Hamiltonian numerically. In doing so, we can identify the steady state density and calculate relevant thermodynamic quantities. We show that the numerical results recover analytical solutions when available. We further identify the second order correction to the populations using analytical analysis as well as the solutions from hierarchical quantum master equation (HQME) \cite{PhysRevB.101.075422,dou2020universal,jin2008exact,tanimura2020numerically}. 


We organize the manuscript as follows. In $\ref{theory}$, we present our analysis as well as our numerical methods to calculate thermodynamic quantities for the resonant level model. In $\ref{results}$, we show our numerical results in non-adiabatic corrections to population, work as well entropy production. We conclude our work in $\ref{con}$.

\section{Theory} \label{theory}
 
\subsection{The driven resonant-level model} \label{resonant}
The driven resonant-level model is probably the simplest yet a heuristic model for the study of quantum thermodynamics in the strongly coupled regime. The model consists of a system, two baths, as well as the couplings between them:  
\beq
\hat{H}=\hat{H}_{\text{sys}}+\hat{H}_{\text{bath}}+\hat{H}_{\text{int}} ,
\eeq
The system consists of one level with a time-dependent on-site energy:  
\beqa
\hat{H}_{\text{sys}} &=&\e_{d}(t)\hat{d}^{\dagger}\hat{d} ,
\eeqa
The bath consists of a set of non-interacting Fermions: 
\beqa
\hat{H}_{\text{bath}}&=&\sum_{k\a}\e_{k\a}\hat{c}_{k\a}^{\dagger}\hat{c}_{k\a} ,
\eeqa
Here $\a =  L$ (or $R$) indicates the left (or right) bath. The couplings between the system and the baths are bilinear: 
\beqa
\hat{H}_{\text{int}}&=&\sum_{k\a}V_{k\a}(\hat{c}_{k\a}^{\dagger}\hat{d}+\hat{d}^{\dagger}\hat{c}_{k\a} ) .
\eeqa
 To describe the strength of the coupling, we define the hybridization function $\Gamma_{\a}(\e)$
\beq
\Gamma(\e)=\sum_{\a}\Gamma_{\a}(\e)=\sum_{k\a}2\pi|V_{k\a}|^{2}\d(\e-\e_{k\a}) \label{ga:mma} .
\eeq
Below, we consider the wide-band limit, such that $\Gamma$ is a constant (independent of energy $\epsilon$). 

\begin{figure}
\includegraphics[scale=0.5]{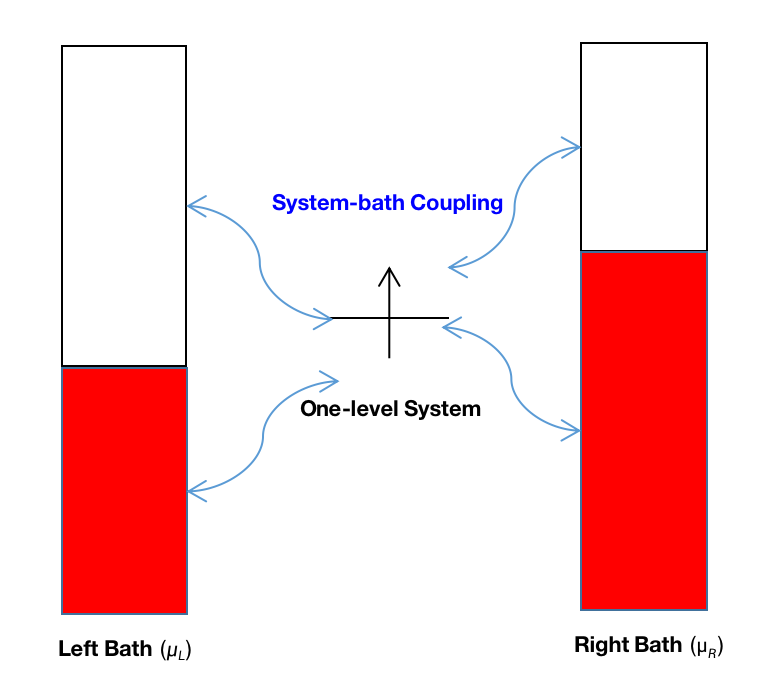}
\caption{ An illustrative picture of the driving resonant-level model.}
\end{figure}

\subsection{Analytical analysis} \label{subsec:analytical}
We now try to define the quantum thermodynamic quantities for the driven resonant-level model. The main quantities of interests are dot population, work, and entropy production:
\begin{eqnarray}
N =  tr ( \hat \rho  \hat{d}^{\dagger}\hat{d}  ) \\
W =  tr ( \hat \rho \partial_t \hat H ) \\
S =  - k_{B}  tr (\hat \rho \ln \hat \rho) 
\end{eqnarray}
Here $\hat \rho$ is the density operator. Note that, we have avoided defined heat and/or energy as they are extensive quantities and requires spectral  treatment.\cite{PhysRevB.98.134306,dou2020universal} Since the Hamiltonian is quadratic, we can rewrite the above quantities using the one-body density operator $\hat \sigma$: 
\begin{eqnarray}
N &=&  \langle d |  \hat \sigma | d \rangle  \\
W &=& \dot \epsilon_d   \langle d |  \hat \sigma | d \rangle  = \dot \epsilon_d  N \\
S &=&  - \text{Tr} ( \hat \sigma \ln \hat \sigma ) - \text{Tr} ( (1-\hat \sigma) \ln (1-\hat \sigma)  ) \label{eq:Stotal}
\end{eqnarray}
Here $\text{Tr}$ denotes the one-body trace (trace in orbital representation). Note that for the resonant level model, work $W$  is proportional to dot population by a factor of $\dot \epsilon_d$, where $\dot \epsilon_d$ is the speed of driving. Such that we will mainly focus on population and entropy below. 

The task is then to calculate the density operator, which can be very difficult for the time-dependent Hamiltonian. Nevertheless, we have the equation of motion for the density matrix
\begin{eqnarray}
\partial_t \hat \sigma = \dot \e_d \partial_{\e_d} \hat \sigma - \frac{i}{\hbar} [ \hat h, \hat \sigma ] 
\end{eqnarray}
Here $\hat h$ is the one body Hamiltonian. In the limit of slow driving, we can expand the density operator into a series of power in the driving speed, 
\beq
\hat \sigma=\hat \sigma^{(0)}+\hat \sigma^{(1)}+\hat \sigma^{(2)}+\cdots ,
\eeq
By matching the order in driving speed from both sides of the above equation, we arrive at a set of equations: 
\begin{eqnarray}
\partial_t \hat \sigma^{(0)} &=& - \frac{i}{\hbar} [ \hat h, \hat \sigma^{(0)} ], \label{0th} \\
\partial_t \hat \sigma^{(n)}  &=& \dot \e_d \partial_{\e_d} \sigma^{(n-1)}  - \frac{i}{\hbar} [ \hat h, \sigma^{(n)}  ], n \ge 1 
\end{eqnarray}
The steady state solution of Eq. \ref{0th} gives us the $0$th order density matrix, $\hat \sigma^{(0)}=\hat \sigma_{ss}$, such that we can proceed to calculate the higher order density matrix
\beq
\begin{aligned}
\hat {\sigma}^{(n)}&=-\dot \epsilon_d \int_{0}^{t}e^{-i\hat{h}(t-t')/{\hbar}}  \partial_{\e_d} \hat{\sigma}^{(n-1)}e^{{i\hat{h}(t-t')}/{\hbar}}dt' \\
&\approx- \dot \epsilon_d \int_{0}^{\infty}e^{-{i\hat{h}t}/{\hbar}}  \partial_{\e_d} \hat {\sigma}^{(n-1)}e^{{i\hat{h}t}/{\hbar}}dt \label{mark} .
\end{aligned} 
\eeq
In the second line of the above equation, we have used the Markovian approximation, assuming the timescale of driving is much smaller as compared with the timescale for the relaxation. The Markovian approximation is consistent with the slow driving approximation.

\subsubsection{Zeroth Order Quantities}
Since the zeroth order entropy is an extensive quantity, we mainly calculate the zeroth order dot population here. To do so, we need to diagonalize the one-body Hamiltonian: 
\beq
\hat{h} =\sum_{k\a}\e_{k\a}  | {\psi}_{k\a} \rangle \langle {\psi}_{k\a} | ,
\eeq
The diagonalization can be done analytically through the following transformation: 
\beq
| {\psi}_{k\a} \rangle =  | {c}_{k\a} \rangle  +  \frac{V_{k\a}}{\e_{k\a}-\e_{d}+i\Gamma/2} \left(   | d \rangle + \sum_{k'\a'}\frac{V_{k'\a'}} {\e_{k\a}-\e_{k'\a'}+i\eta)}  | {c}_{k'\a'} \rangle \right)     \label{di:ag} .
\eeq
With the diagonalized Hamiltonian, we can determine the steady-state single particle density matrix
\begin{eqnarray}
\hat \sigma^{(0)} = \sum_{k\a} f (\e_{k\a} - \mu_\alpha )  | {\psi}_{k\a} \rangle \langle {\psi}_{k\a} | 
\end{eqnarray}
Here, $f $ is the Fermi function $f(\e_{k\a} - \mu_\alpha )=\left[1+e^{\b(\e_{k\a}-\mu_{\a})} \right]^{-1}$, and $\mu_\alpha$ is the chemical potential for the $\alpha$ lead.

With the analytical results shown above, we can obtain the zeroth order population:  
\beq
\begin{aligned}
N^{(0)}=\left<d\right|\hat \sigma^{(0)}\left|d\right> = \frac{1}{2\pi}\int d\e A(\e)\bar{f}(\e) ,
\end{aligned} \label{N0}
\eeq
Here $A(\e)$ is the spectral function
\beqa
A(\e)=\frac{\Gamma}{(\e-\e_{d})^{2}+(\Gamma/2)^{2}},  \label{A}
\eeqa
and $\bar{f}(\e)$ is the weighted Fermi function $\bar{f}(\e)=\sum_{\a}\frac{\Gamma_{\a}}{\Gamma}\left[1+e^{\b(\e-\mu_{\a})} \right]^{-1} $. Details of the deviation are shown in Appendix $\ref{appA}$. 

\subsubsection{First Order Quantities}
Now we turn to the first order correction. With $0$th order $\hat \sigma^{(1)}$, we can proceed to calculate the first order density: 
\beq
\hat \sigma^{(1)}=- \dot \epsilon_d \int_{0}^{\infty}e^{-{i\hat{h}t}/{\hbar}}  \partial_{\e_d} \hat {\sigma}^{(0)}e^{{i\hat{h}t}/{\hbar}}dt. \label{eq:sigma1}
\eeq
The first order correction to the population is then given by 
\beq
N^{(1)} =\left<d\right| \hat \sigma^{(1)}\left|d\right>=-\frac{\hbar\dot{\e}_{d}}{4\pi}\int d\e A^{2}{\partial}_{\e}\bar{f}(\e) \label{N1}.
\eeq
Detailed derivation can be found in the appendix of Refs. \citen{PhysRevB.98.134306,PhysRevLett.119.046001}. Note that the population is linearly proportional to driving speed $\dot \epsilon_d$. Such that the first order work depends on $\dot \epsilon_d$ quadratically, $W^{(1)} =\gamma^{(1)} \dot \epsilon^2_d$. Here $\gamma^{(1)}$ is the frictional coefficient: 
\beq
\gamma^{(1)}=\frac{N^{(1)}}{\dot{\e}_{d}}=-\frac{\hbar}{4\pi}\int d\e A^{2}{\partial}_{\e}\bar{f}(\e) \label{gam1}, 
\eeq
which accounts for the dissipative effects. 

We could also define a correlation function of the random force to qualify the fluctuation
\begin{eqnarray}
D_{\mu\nu}&=&\int_0^\infty \langle e^{{i\hat{H}t}/{\hbar}}  \d\hat{F}_{\mu}  e^{-{i\hat{H}t}/{\hbar}}  \d\hat{F}_{\nu}(0) \rangle_S \label{eq:Corr}\\
\d\hat{F}_{\a} &=&{\partial}_{\a}\hat{H} - tr({\partial}_{\a}\hat{H}\hat{\rho}_{ss})  \label{eq:randomF}
\end{eqnarray}
$\langle \cdots \rangle_S$ denote the symmetric average of the correlation functions. Here $\alpha$ and $\mu$ are degrees of freedom for the motion. In the driven resonant level model (with $\alpha = \mu = \epsilon_d $), we can calculate the fluctuation analytically
\beq
D=\frac{\hbar}{4\pi}\int d\e A^{2}\bar{f}(\e)(1-\bar{f}(\e)) \label{D1} .
\eeq

We then turn to the first order correction to entropy, which can be calculated through Taylor expansion of Eq. \ref{eq:Stotal}: 
\beq
S^{(1)}=-k_{B} \text{Tr} \left[\hat \sigma^{(1)}\ln \hat \sigma^{(0)} - \hat \sigma^{(1)}\ln{(1 - \hat \sigma^{(0)})}\right] ,
\eeq
The non-adiabatic entropy rate is the derivative of the above equation. 
\beq
\begin{aligned}
\dot{S}^{(2)}=\partial_t S^{(1)} = \frac{\dot{Q}^{(2)}}{T}+\d \dot{S}, \label{eq:dotS2}
\end{aligned}
\eeq
The first term in Eq. \ref{eq:dotS2} is the first order correction to the heat \cite{dou2020universal} 
\beq
\dot{Q}^{(2)}=-k_{B} T  \dot{\e}_{d} \text{Tr} (\left[{\partial}_{\e_{d}}{\sigma}^{(1)}\ln{{\sigma}^{(0)}}-{\partial}_{\e_{d}}{\sigma}^{(1)}\ln{(1 -{\sigma}^{(0)})}\right] .
\eeq
The second term $\d \dot{S}$ in Eq. \ref{eq:dotS2} is the additional entropy production rate due to the external driving
 \beq
 \begin{aligned}
 \d\dot{S}= &-k_{B}\dot{\e}_{d}\left[\hat {\sigma}^{(1)}{\partial}_{\e_{d}}\ln\hat {\sigma}^{(0)} -\hat \sigma^{(1)} {\partial}_{\e_{d}}\ln{(1-\hat {\sigma}^{(0)})} \right] 
 \end{aligned}
 \eeq

At equilibrium, we can calculate the entropy production rate explicitly and show that 
\beq
\gamma^{(1)}=\beta D =\frac{\hbar T \d\dot{S}}{\dot{\e}_{d}} \label{eq:relation}.
\eeq
The first equality in the above equation is the fluctuation-dissipation theorem. Under non-equilibrium condition, the fluctuation-dissipation theorem is no longer obeyed and we have not been able to determine the entropy production rate. In Sec. \ref{subsec:numerical}, we will show how to evaluate entropy production rate numerically. 
 
\subsubsection{Second Order Quantities}
Evaluation of the second order corrections to thermodynamic quantities is very tricky, which will require 
\beq
\hat \sigma^{(2)}=- \dot{\e}_{d} \int_{0}^{\infty}e^{-{i\hat{h}t}/{\hbar}}{\partial}_{\e_d} {\sigma}^{(1)}e^{{i\hat{h}t}/{\hbar}}dt .
\eeq
Nevertheless, as shown in Appendix \ref{app:B}, we manage to evaluate the second order correction to the population analytically for the equilibrium case, 
\beq
N^{(2)}=\left<d\right|\hat \sigma^{(2)}\left|d\right>=\frac{\hbar^{2}\dot{\e}_{d}^{2}}{12\pi}\int d\e A^{3}{\partial}_{\e}^{2} {f}(\e) \label{N2} .
\eeq
Similarly to the first order case, we can define the friction term $\gamma^{(2)}$ as 
\beq
\gamma^{(2)}=\frac{N^{(2)}}{\dot{\e}_{d}^{2}}=\frac{\hbar^{2}}{12\pi}\int d\e A^{3}{\partial}_{\e}^{2}{f}(\e) \label{gam2} .
\eeq
Out of equilibrium, we do not have analytical results for the second order corrections. We could simply replace $f$ by $\bar f$ in the above equation as our trial results. In general, however, the higher order correction can be evaluated numerically using hierarchical quantum master equation (HQME).

\subsection{Numerical Methods} \label{subsec:numerical} 
The derivations in evaluating the non-adiabatic corrections can be very lengthy and a compact form of solutions may not available for certain quantities. In this subsection, we consider a finite number of degrees of freedom for the baths in a similar fashion to the Ref. \citen{oz2019numerical}, such that we can evaluate non-adiabatic corrections numerically. 

To do so, we discrete the baths degrees of freedom, and we build the one-body Hamiltonian $\hat h$ with finite number of levels from the baths. $\hat h$ can be written in the matrix form: 
\begin{eqnarray}
\hat h = \begin{pmatrix}
\e_d & \bold{V_L}&  \bold{V_R}\\
\bold{V_L}^{\dagger} & \hat \epsilon_L &0  \\
\bold{V_R}^{\dagger} &0& \hat \epsilon_R
\end{pmatrix}
\end{eqnarray}
Here, $ \bold{ V_\alpha} = \{ V_{k\alpha} \} $ is a vector representation of the system-bath coupling, and $\hat \e_\alpha =  \{ \epsilon_{k\alpha} \}$ is a matrix representation of the energies in the baths. 

We then proceed to build the steady state density matrix $\hat \sigma_{ss}$. We first define non-interacting density operator
\begin{eqnarray}
\hat \sigma_{0} = \begin{pmatrix}
f(\e_d - \frac12 \mu_L - \frac12  \mu_R)  &0& 0\\
0& f (\hat \e_L -\mu_L)  &0\\
0& 0& f(\hat \e_R -\mu_R)
\end{pmatrix}
\end{eqnarray}
The interacting steady-state density operator is obtained by transforming  $\hat \sigma_{0}$ in the basis where $\hat h$ is diagonal and zero out all non-diagonal terms of $U^\dagger  \hat \sigma_{0} U$. Here $U^\dagger$ diagonalizes the $\hat h$. Such that $\hat \sigma_{ss}$ and $\hat h$ are both diagonal in such a basis. 
\begin{eqnarray}
\hat h = \sum_m \epsilon_m | m \rangle \langle m | \\
\hat \sigma_{ss}  = \sum_m \sigma_m | m \rangle \langle m | \label{eq:rho_ssNum} 
\end{eqnarray}  
Such a scheme is similar to the scattering formulation of the density operator. \cite{ness2017nonequilibrium} 

With the eigenvalues and eigenbasis of $\hat h$ and $\hat \sigma_{ss}$, the zeroth order population is easily obtained:  
\beq
N^{(0)}= \sum_m | \langle d |m \rangle|^2 \sigma_m 
\eeq
The first order correction requires $\hat \sigma^{(1)}$ (Eq. \ref{eq:sigma1}), which can be evaluated in the eigenbasis of $\hat h$: 
\begin{eqnarray}
\hat \sigma^{(1)} = \pi\hbar\dot{\e}_{d}\sum_{mn} |m \rangle  \langle m | \partial_{\e_d} \sigma_{ss} |n \rangle   \d(\e_m -\e_n) \langle n | 
\end{eqnarray}
Using Hellman-Feynman theorem, we can rewrite the above as (see the appendix in Ref. \citen{PhysRevB.98.134306}) 
\begin{eqnarray}
\hat \sigma^{(1)} = \pi\hbar\dot{\e}_{d}\sum_{mn} |m \rangle \left<m |d\right> \frac{\sigma_m - \sigma_n}{\e_m -\e_n}  \d(\e_m -\e_n) \left<d|n \right>\langle n | 
\end{eqnarray}
Such that the first order population is 
\beq
N^{(1)}= \pi\hbar\dot{\e}_{d}\sum_{mn}  |\left<m |d\right>|^2 \frac{\sigma_m - \sigma_n}{\e_m -\e_n}  \d(\e_m -\e_n) |\left<d|n \right> |^2  \label{eq:N1num}
\eeq
In numerical simulations, we replace the delta function by a Gaussian with a broadening parameter $\eta$: 
\begin{eqnarray}
\pi \d(\e_m-\e_n) &\rightarrow& \sqrt{\frac{\pi}{2t^{2}}}\exp{\left[\frac{(\e_m-\e_n)^{2}}{2\eta^{2}}\right]} \label{gauss} 
\end{eqnarray}
We also take care of the denominator $\e_m -\e_n$ in Eq. \ref{eq:N1num} as
\begin{eqnarray}
\frac{\sigma_m - \sigma_n}{\e_m -\e_n}  &\rightarrow&  \Re \frac{\sigma_m - \sigma_n}{\e_m -\e_n + i\eta} 
\end{eqnarray}
In practice, the broadening parameter $\eta$ is set to be in the order of the energy spacing to better converge the results. 

The correlation function in Eq. \ref{eq:Corr} can be recast in the eigenbasis (see the supplementary material in Ref. \citen{PhysRevLett.119.046001}): 
\beq
D = \pi\hbar\dot{\e}_{d}\sum_{mn}  |\left<m |d\right>|^2 \sigma_m (1-\sigma_n) \d(\e_m -\e_n) |\left<d|n \right> |^2 
\eeq
which is readily to be calculated. Finally, $\delta \dot S$ can be written in the eigenbasis as 
\beq
\delta \dot S = \pi\hbar\dot{\e}_{d}\sum_{mn}   \langle m | \hat \sigma^{(1)} |n \rangle   \langle n |  \partial_{\e_d} (\ln \hat \sigma_{ss} - \ln (1-\hat \sigma_{ss}) ) |m \rangle
\eeq
Here, $\langle m | \partial_{\e_d} \ln \hat \sigma_{ss} |n \rangle  $ is done numerically with finite difference. Note that $|m\rangle$ is the adiabatic basis. To calculate the matrix element properly, we rotate the matrix $\ln \hat \sigma_{ss}$ back to the diabatic basis. After done derivative, we rotate the matrix back to the adiabatic basis.

\section{Results and discussion} \label{results}

\subsection{Weak coupling limit: Master Equations}
In this subsection, we consider the the weak coupling limit. In such a limit, we can trace out the bath degrees of freedom and derive an equation of motion for the dot density $\sigma_d$ only \cite{paperII,mulFrictionPaperJCP2016}:  
\beq
{\partial}_{t}\sigma_d = -\dot{\e_d}{\partial}_{\e_d}\sigma_d -\frac{\Gamma}{\hbar}\left[\sigma_d -\bar{f} (\epsilon_d \right]
\label{cme} .
\eeq
Again, we can expand the dot density into a series of powers in terms of the driving speeds, $\sigma_d=\sigma_d^{(0)}+\sigma_d^{(1)}+\sigma_d^{(2)}+\cdots$. Similar to the procedures in Sec.  \ref{subsec:analytical}, we obtain the $n$th order population as
 \beqa
 N^{(0)}=\bar{f}(\e_{d}) ,\\
 N^{(1)}=-\frac{\hbar\dot{\e}_{d}}{\Gamma}{\partial}_{\e_{d}}\bar{f}(\e_{d}) ,\\
 N^{(2)}=\frac{\hbar^{2}\dot{\e}_{d}^{2}}{\Gamma^{2}}{\partial}_{\e_{d}}^{2}\bar{f}(\e_{d}) \label{UN2} .
 \eeqa
Similarly, we can also obtain the friction and fluctuation,  
 \beqa
 \gamma^{(1)}=-\frac{\hbar}{\Gamma}{\partial}_{\e_{d}}\bar{f}(\e_{d}) ,\\
 D=\frac{\hbar k_{B}T}{\Gamma}\bar{f}(\e_{d})[1-\bar{f}(\e_{d})] 
 \eeqa
 To get the entropy production rate, we define the entropy for the dot as $S = - \sigma_d \ln \sigma_d - (1-\sigma_d) \ln (1-\sigma_d) $. Similar to the procedures in \ref{subsec:analytical}, we arrive at the entropy production rate as
  \beqa
  \frac{\hbar T\d \dot{S}}{\dot{\e}_{d}}=[\gamma^{(1)}]^{2}/(\b D) .
  \eeqa
 Note that the above equation is only true for the driven resonant level model in the weak coupling limit. We expect that our numerical results can reduce to the master equation results in the limit of $\Gamma<kT$. What is missing in the master equation is mainly the level broadening effects. Hence, we refer to the master equation results as unbroadened results.

\subsection{Results}
We now present our results in this subsection. In our numerical calculations, number of the levels is set to be $1000$ for both leads, and level spacing is set to be uniform. The bandwidth is $100\Gamma$. We have also set $\hbar \dot{\e}_{d}=0.1\Gamma^{2}$ and the broadening term $\eta$ is equal to the energy spacing.

\subsubsection{Zeroth and First Order Results}
\begin{figure} [h]
\includegraphics[scale=0.6]{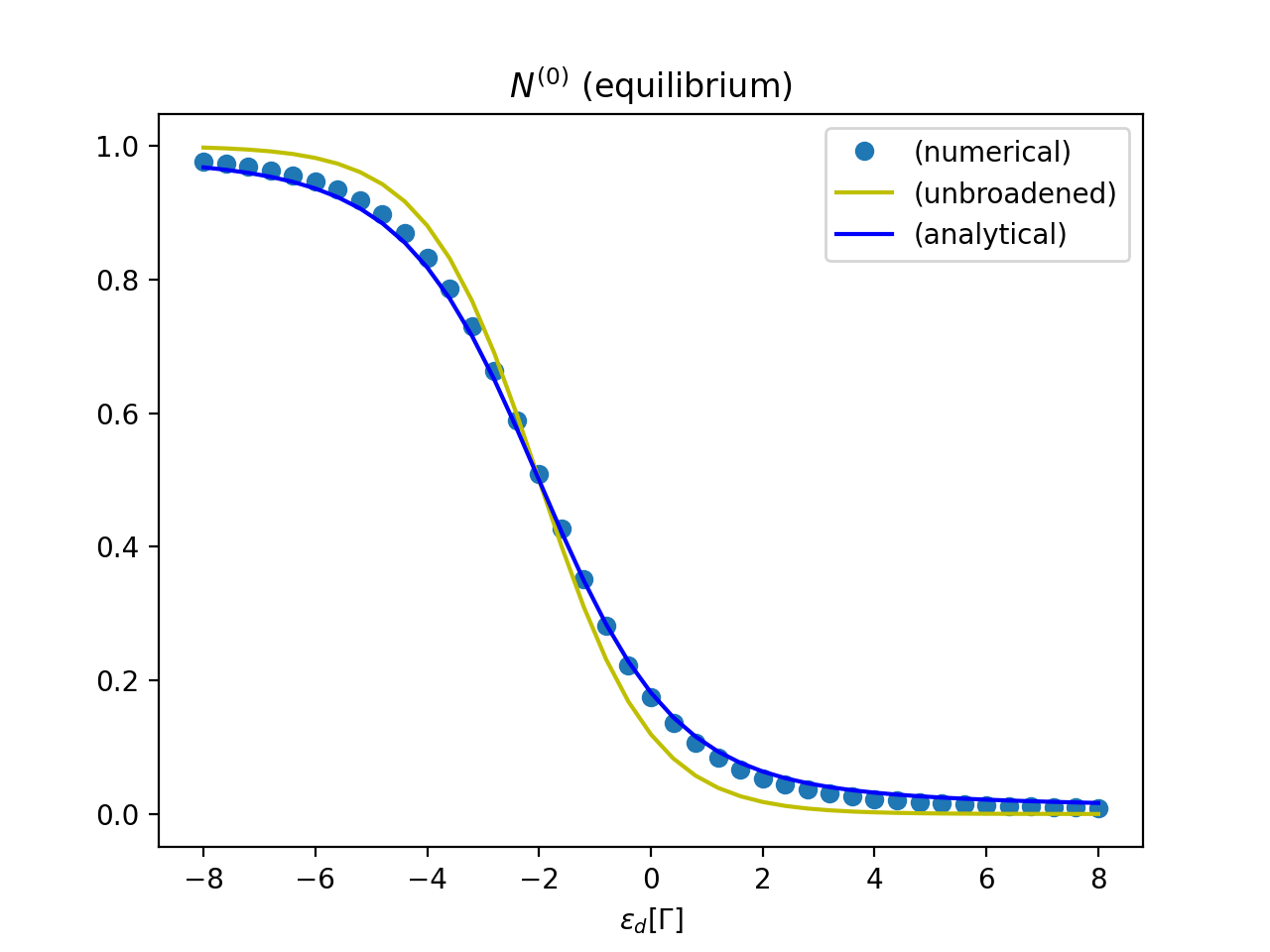}
\includegraphics[scale=0.6]{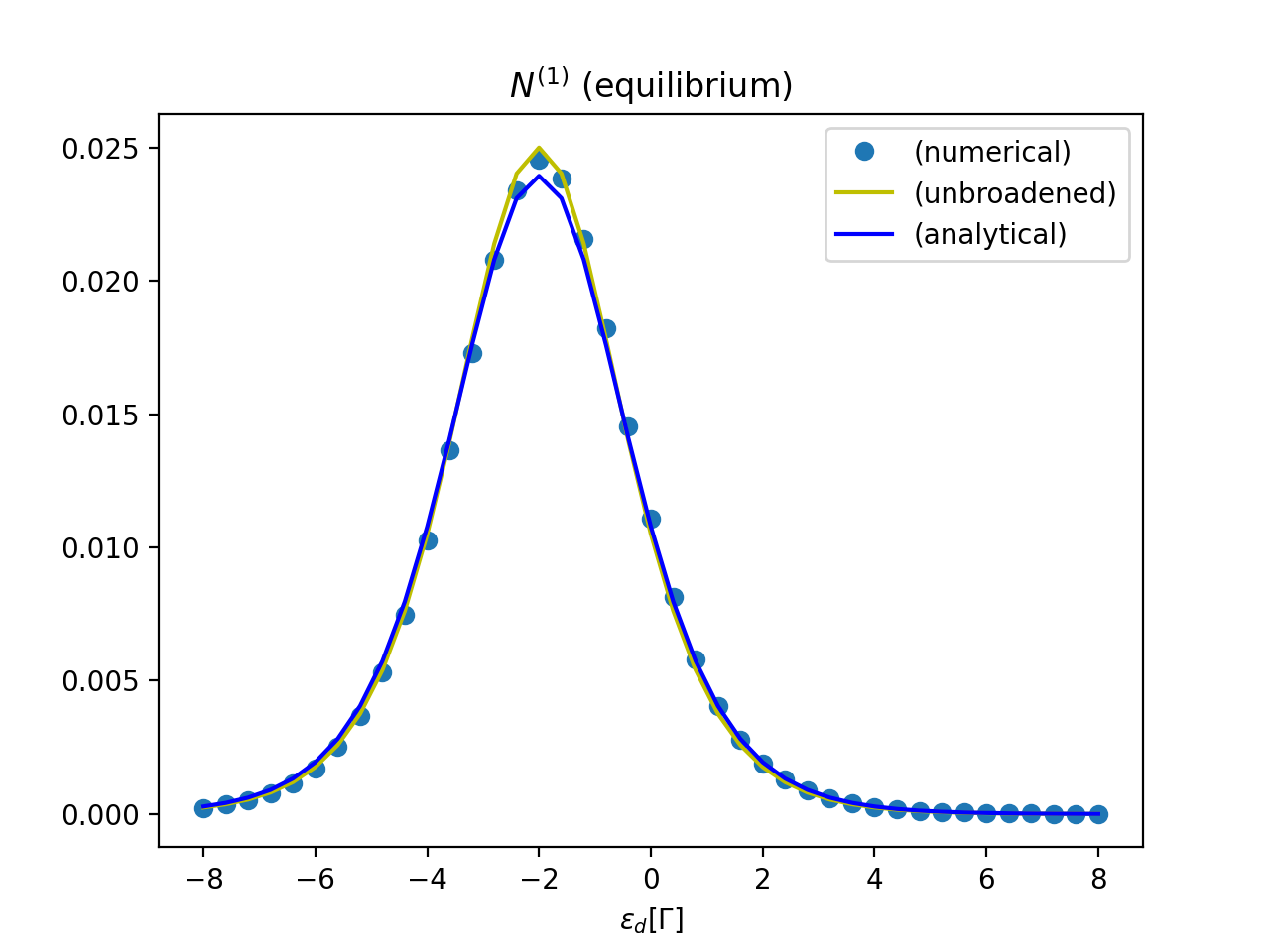}
\caption{ $N^{(0)}$ and $N^{(1)}$ as a function of $\e_d$ at equilibrium. $\mu_L = \mu_R = -2 \Gamma$, $\Gamma = k_B T$. }
\label{fig:1}
\end{figure}

We first look at the zeroth and first order corrections to the thermodynamic quantities. In Fig. \ref{fig:1}, we plot the dot population $N^{(0)}$ and $N^{(1)}$ as a function of $\epsilon_d$ at equilibrium where $\mu_{L}=\mu_{R} = - 2\Gamma$. In such a case, analytical results are available. Notice the good agreement between the numerical results and the analytical results for both the zeroth order and first order populations.  In the numerical calculations, the main source of the error is the discretization of finite bandwidth. This agreement verify the reliability of the numerical method. The results from master equation having a sharper feature as compared to the other results, suggesting the missing of the level broadening effects. Note also that the first order correction to the population $N^{(1)}$ shows a peak at $\epsilon_d = \mu_L = \mu_R$. 

\begin{figure} [h]
\includegraphics[scale=0.8]{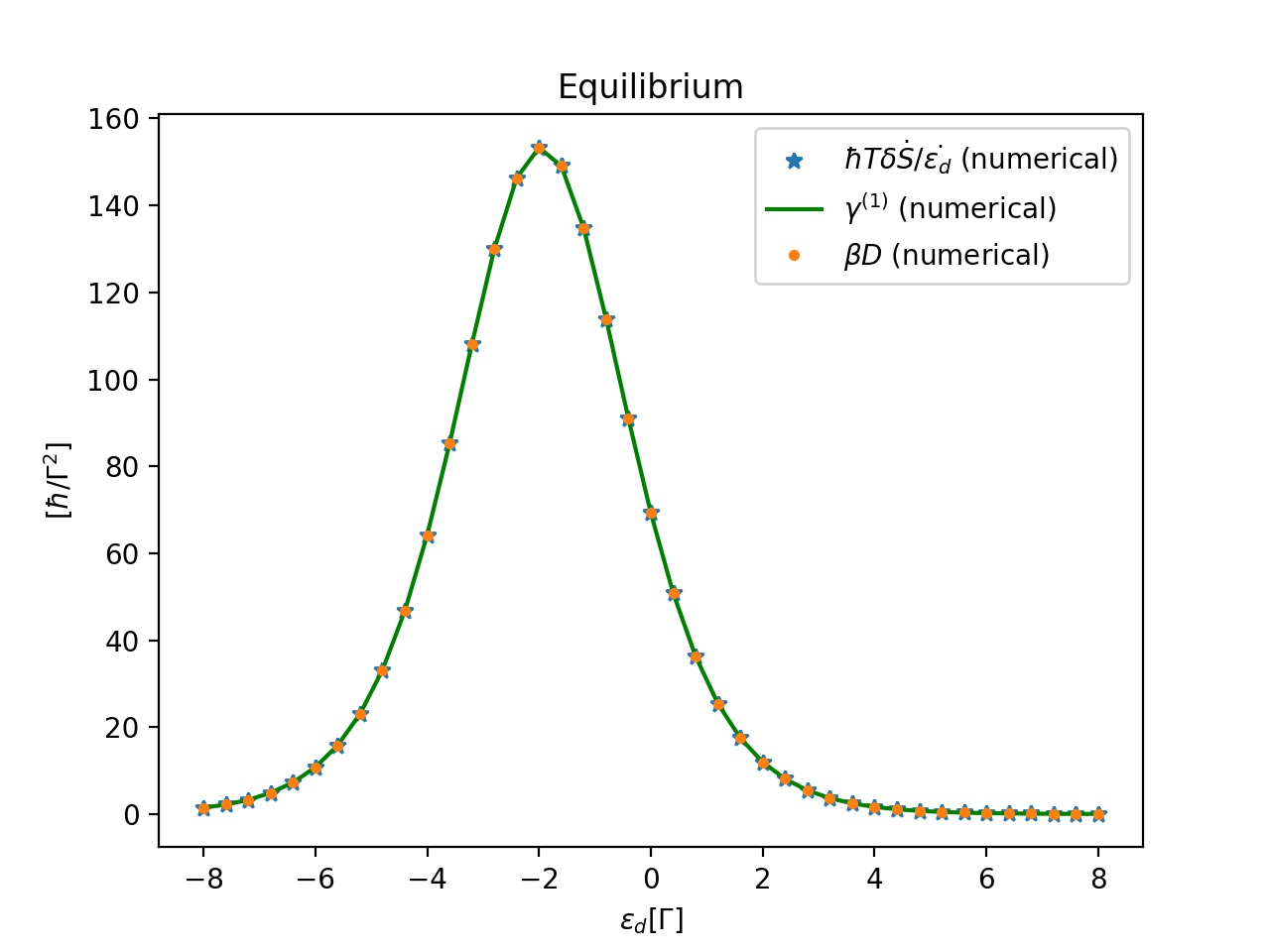}
\caption{Friction, fluctuation, and entropy production rate as a function of $\e_d$ at equilibrium. $\mu_L = \mu_R = -2 \Gamma$, $\Gamma = k_B T$. }
\label{fig:2}
\end{figure}

In Fig. \ref{fig:2}, we plot friction, fluctuation, and entropy production rate as a function of $\e_d$ at equilibrium. In Eq. \ref{eq:relation}, we have shown that friction, fluctuation, and entropy production rate are proportional to each other analytically without non-equilibrium condition. Indeed, our numerical calculations further verify this statement. Such an agreement will no long exist under non-equilibrium condition (as shown below). 

\begin{figure} [h]
\includegraphics[scale=0.6]{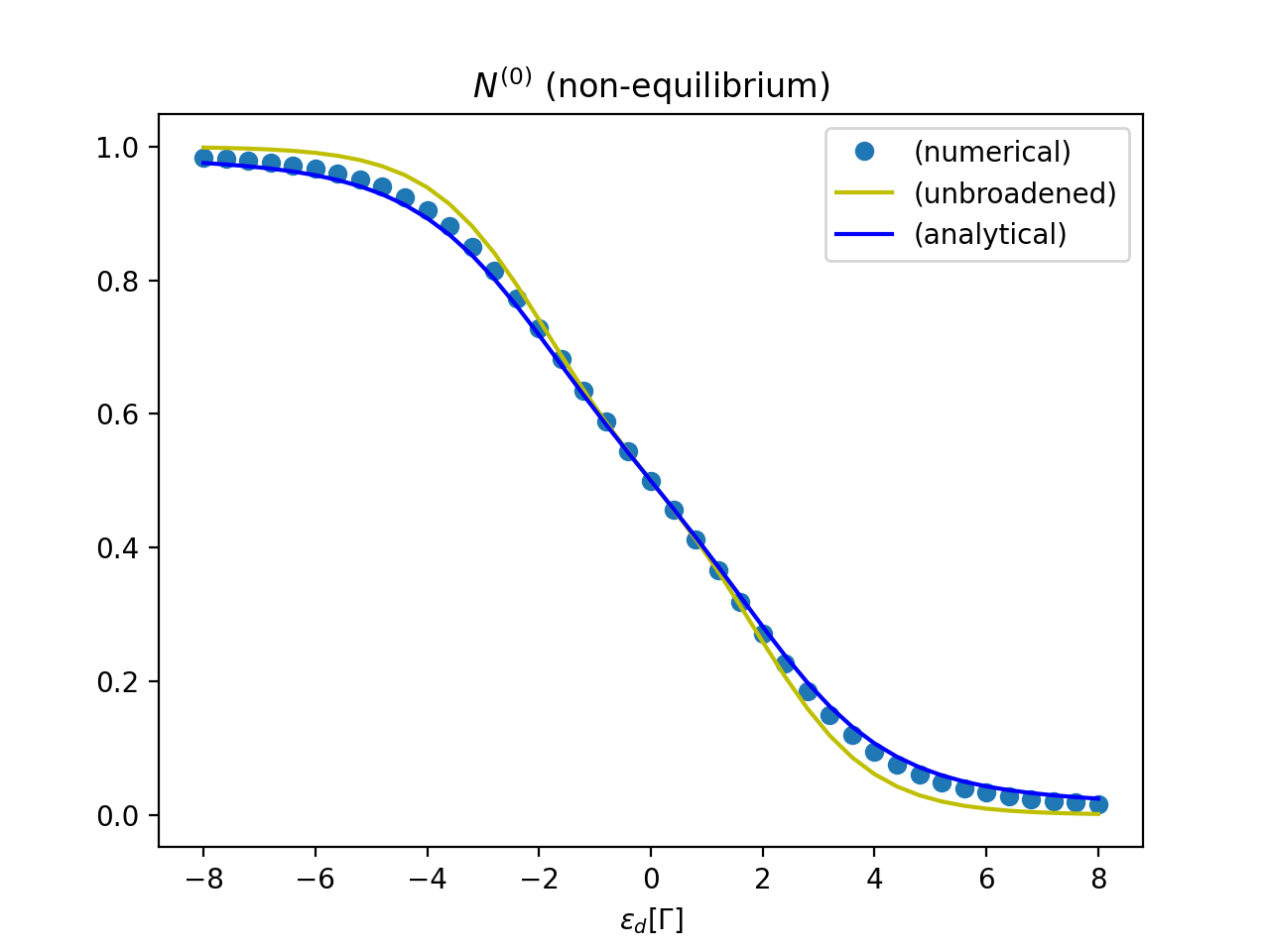}
\includegraphics[scale=0.6]{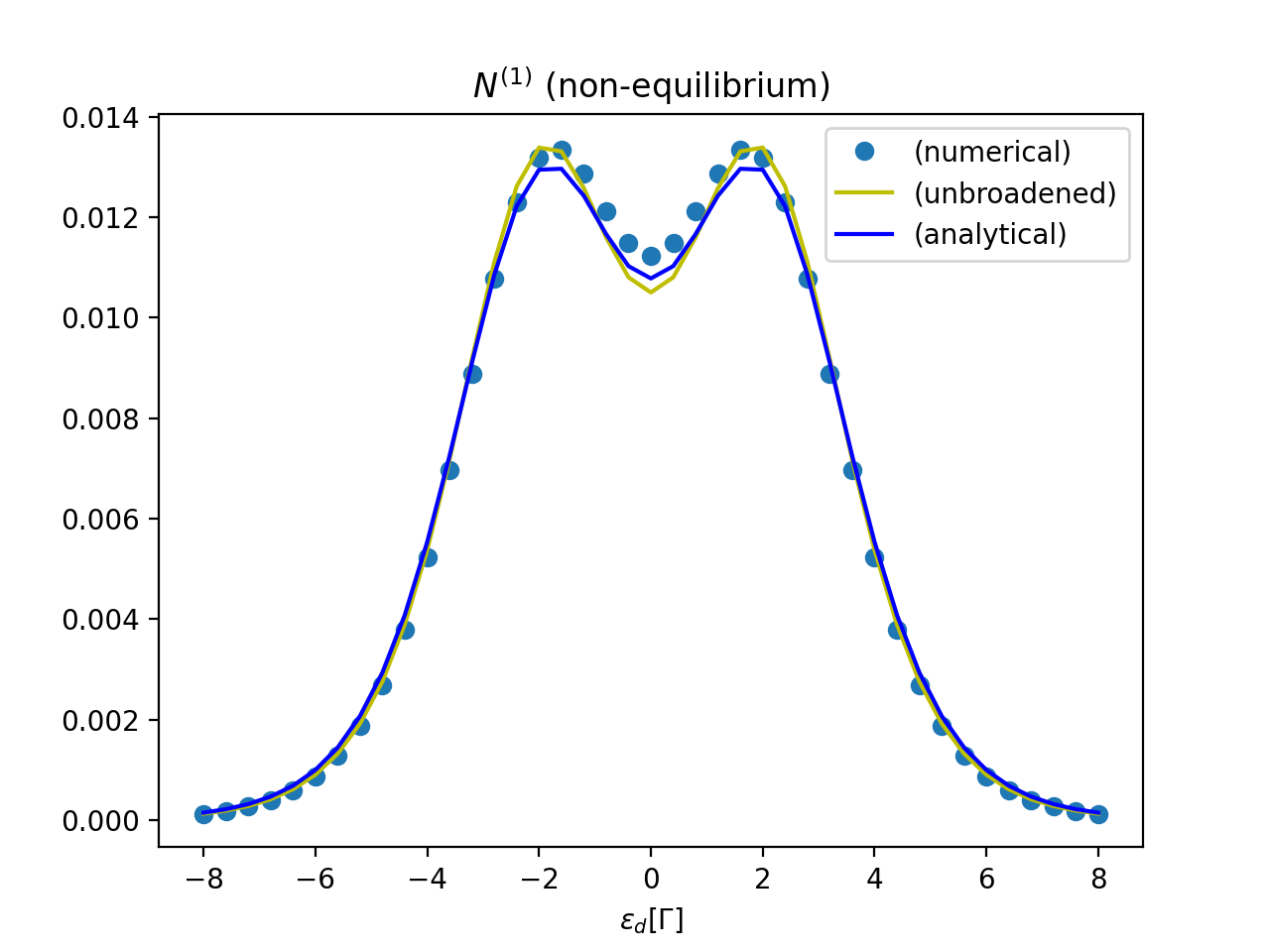}
\caption{$N^{(0)}$ and $N^{(1)}$ as a function of $\e_d$ under nonequilibrium condition. $\mu_L = -\mu_R = -2 \Gamma$, $\Gamma = k_B T$. }
\label{fig:3} 
\end{figure}

Now, let us look at the non-equilibrium case. In Fig. \ref{fig:3}, we plot $N^{(0)}$ and $N^{(1)}$ as a function of $\e_d$ under non-equilibrium condition, where $\mu_{L}=-\mu_{R}=-2\Gamma$. Again, we see good agreement between the analytical solution and the numerical calculation. The first order population $N^{(1)}$ shows two peaks at $e_d = \mu_{L}$ and $e_d=\mu_{R}$ respectively, resulting a dip at $\e_d = \frac12 (\mu_L + \mu_R)$. Hence, the peaks are Fermi resonance in nature. Again, results from quantum master equation show shaper dips and peaks due to lack of broadening effects. 

\begin{figure} [h] 
\includegraphics[scale=0.8]{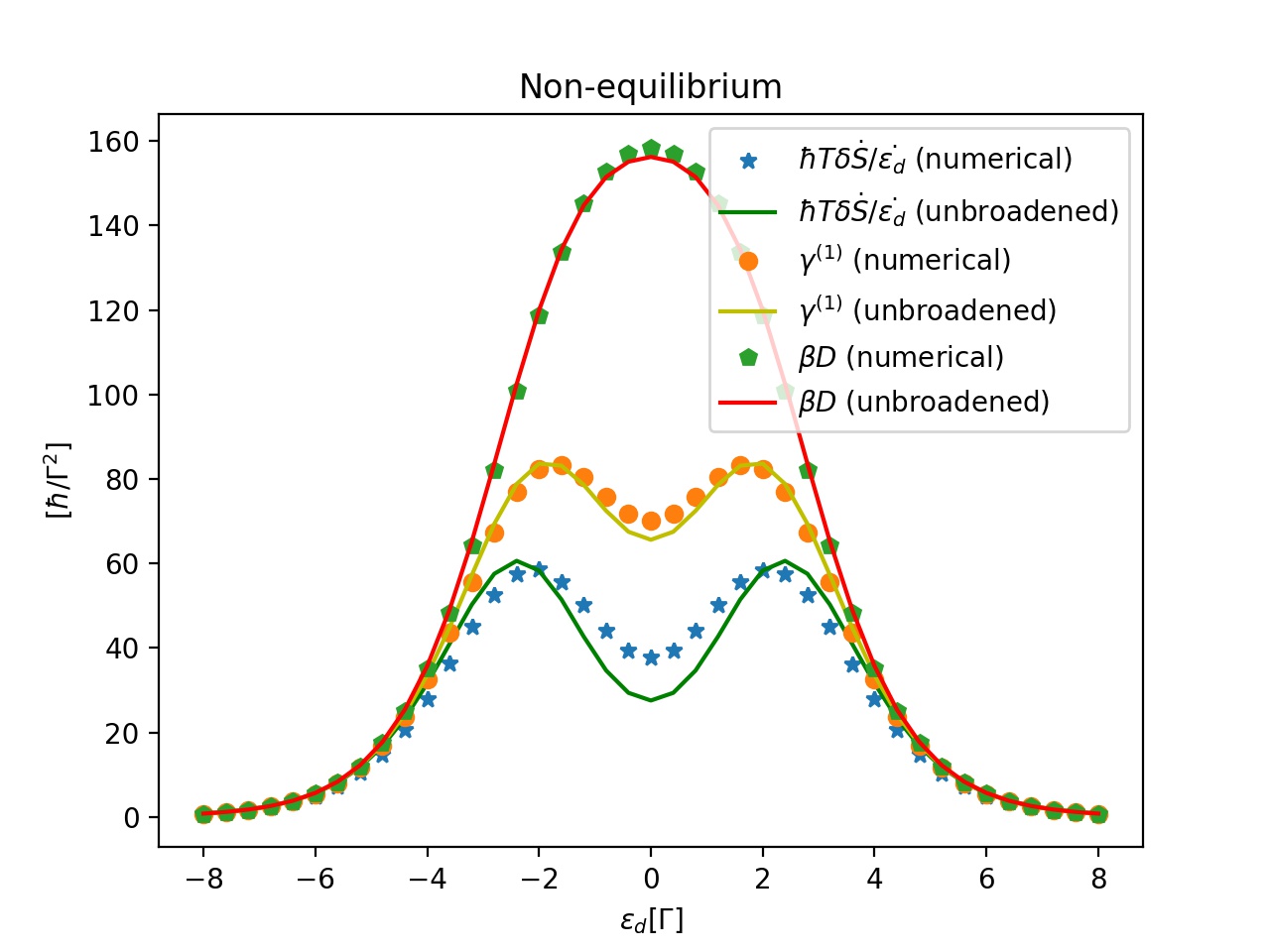}
\caption{Friction, fluctuation, and entropy production rate as a function of $\e_d$ under nonequilibrium condition. $\mu_L = -\mu_R = -2 \Gamma$, $\Gamma = k_B T$.}
\label{fig:4}
\end{figure}

In Fig. \ref{fig:4}, we plot friction, fluctuation, and entropy production rate as a function of $\e_d$ under non-equilibrium condition. Now we see that, unlike the equilibrium case, friction, fluctuation, and entropy production rate do not agree with each other. In fact, in general all three quantities can be recast into correlation functions: \cite{dou2020universal} 
\begin{eqnarray}
\delta \dot S &=&  {k_B}{\bar \beta^2} \sum_{\alpha \nu} \dot R_{\alpha} \dot R_{\nu}   \int_0^\infty  \langle \delta \hat{\mathcal{F}}_\alpha (t)  \delta \hat{\mathcal{F}}_\nu \rangle_{K} dt \\
\gamma_{\alpha \nu}  &=&  \bar \beta \int_0^\infty  \langle \delta \hat{{F}}_\alpha (t)  \delta \hat{\mathcal{F}}_\nu \rangle_{K} dt \\
D_{\alpha \nu}  &=&   \int_0^\infty  \langle \delta \hat{{F}}_\alpha (t)  \delta \hat{{F}}_\nu \rangle_S dt .
\end{eqnarray}
Here $\langle \cdots \rangle_K$ denotes Kubo transformed average of the correlation functions. $\bar \beta$ is the reduced temperature \cite{ness2017nonequilibrium}.  $\delta \hat{\mathcal{F}}_\alpha $ can be seen as the generalized random force: 
\begin{eqnarray}
\delta \hat{\mathcal{F}}_\alpha = - \frac{1}{\bar \beta}  \partial_\alpha \ln \hat \rho_{ss} = \partial_\alpha \hat H - \partial_\alpha \hat Y - Tr(\hat \rho_{ss}  (\partial_\alpha \hat H -\partial_\alpha \hat Y) ) 
\end{eqnarray}   
Here $\hat Y$ is an operator that accounts for particle transport. \cite{PhysRevLett.70.2134} $\delta \hat{{F}}_\alpha$ is the random force defined in Eq. \ref{eq:randomF}. Hence, friction, fluctuation, and entropy production rate are just correction functions of the two random force operators. Since $\delta \dot S$ and $D_{\alpha \nu}$ are the correlation functions of the same random force, such that $\delta \dot S$ and $D_{\alpha \nu}$ are positive definite, whereas in general $\gamma_{\alpha \nu}$ is not positive definite under non-equilibrium condition. Here, for the resonant level model, we see that $\gamma_{\alpha \nu}$ remains positive even out of equilibrium. However, For more complicated model, this is not true. \cite{beilstein}

\subsubsection{Second Order Results}
\begin{figure}
\includegraphics[scale=0.6]{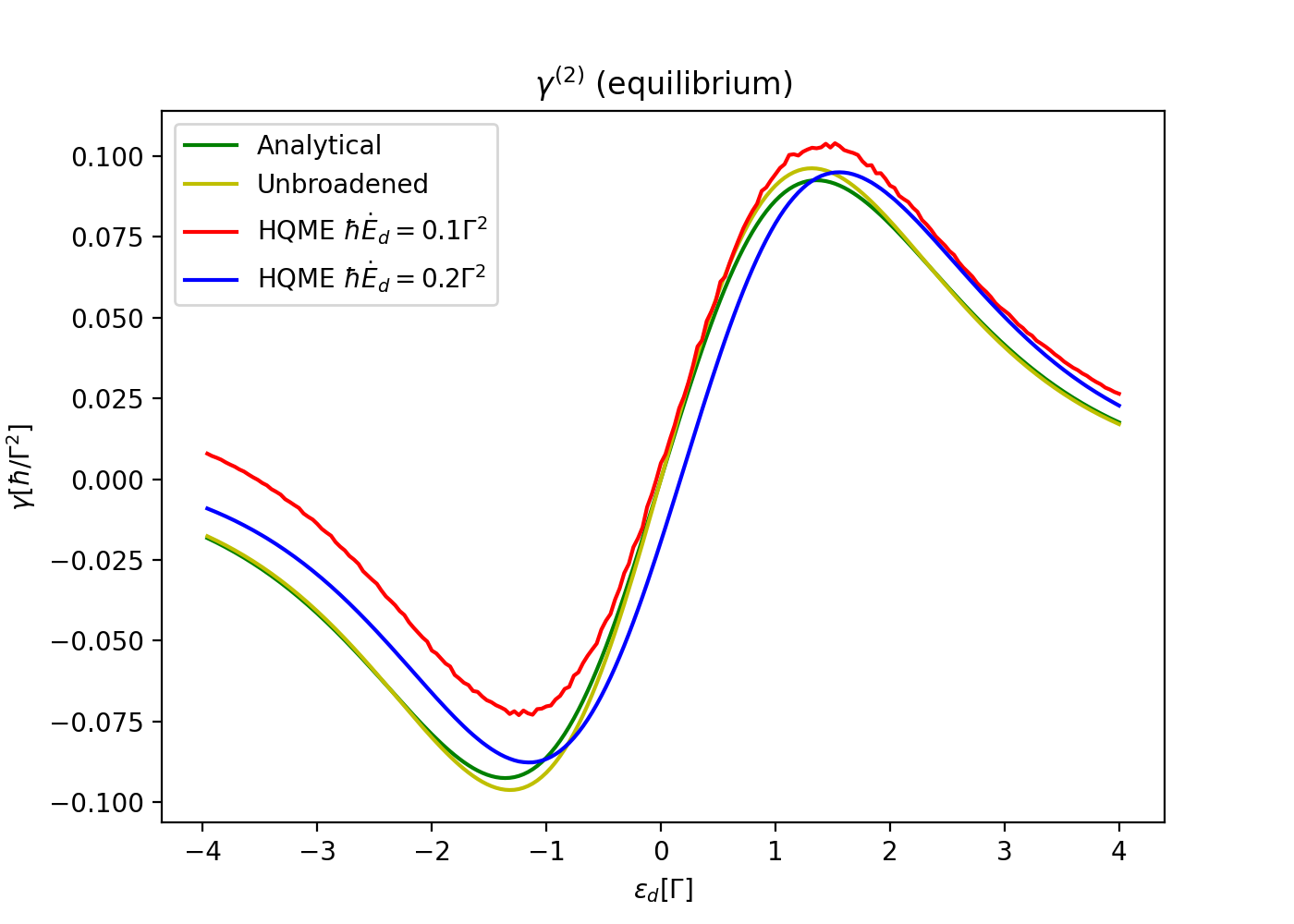}
\includegraphics[scale=0.6]{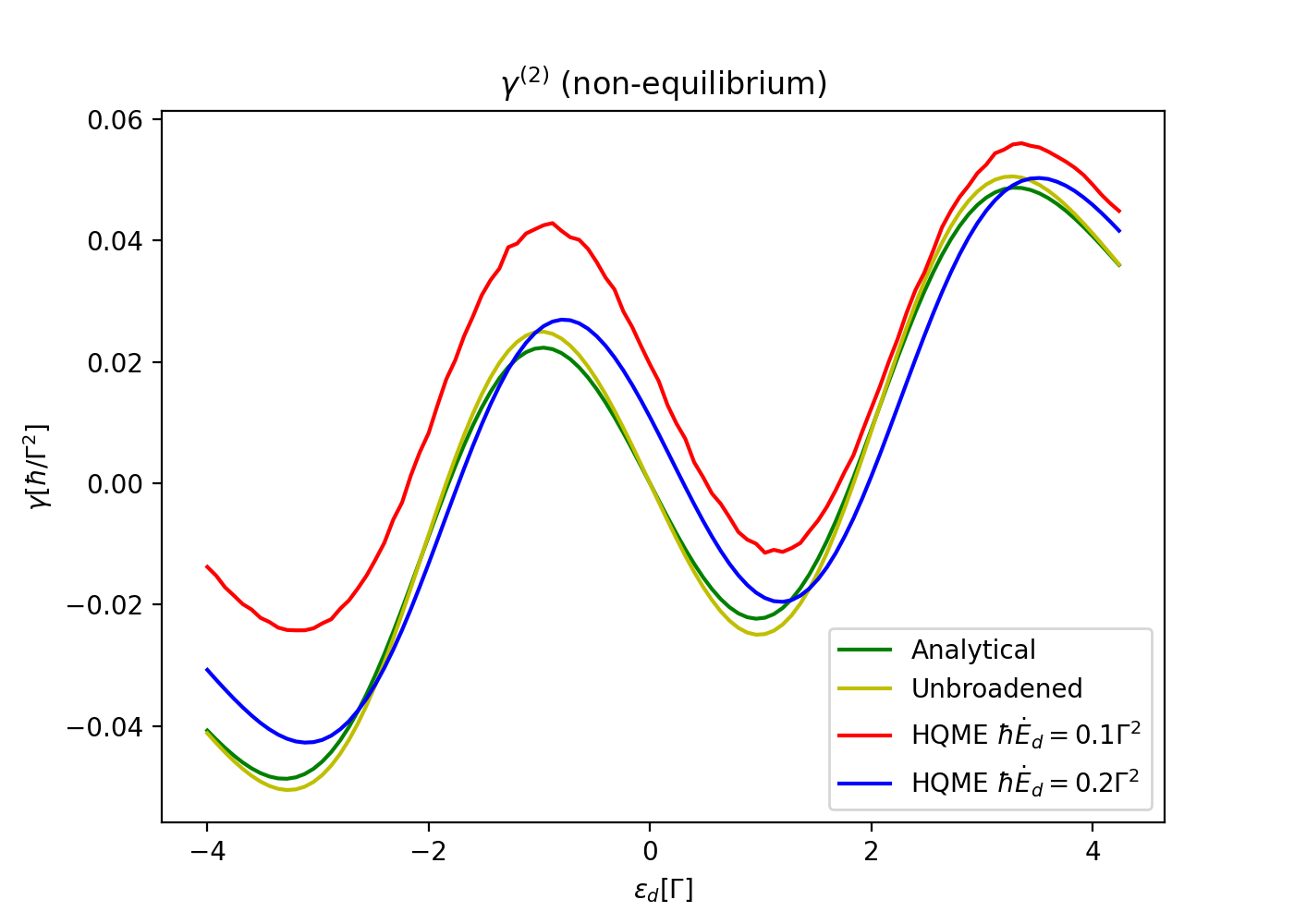}
\caption{ $\gamma^{(2)}$ as a function of $\e_{d}$ under equilibrium ($\mu_L = \mu_R = 0$) and nonequilibrium condition ($\mu_L = -\mu_R = - 2 \Gamma$).  $\Gamma = k_B T$. }
\label{fig:currentU}
\end{figure}

We now look at the second order correction to the population (or second order friction). Here, HQME is used to simulate the exact dynamics for the driven resonant level model, where we can calculate the exact population as a function of time. We then estimate the second order friction by subtracting the zeroth and first order correction: 
\beq
\gamma^{(2)}=\left(N-N^{(0)}-\dot{\e}_{d}\gamma^{(1)}\right)/\dot{\e}_{d}^{2} ,
\eeq
In the above equation, we have ignored the third order and higher correction. In Fig. \ref{fig:currentU}, we plot $\gamma^{(2)}$ as a function of $\e_{d}$ from HQME with different driving speeds. At equilibrium, HQME results reproduce analytical results in Eq. \ref{gam2}. Out of equilibrium, we do not have analytical results for $\gamma^{(2)}$. However, we simply replace $f$ by $\bar f$ in Eq. \ref{gam2} as our trial solution. Such a solution agrees with the HQME results as well as results from master equations. Indeed, future work must verify this trial solution for the second order correction out of equilibrium.

\section{Conclusion} \label{con}
Using the expansion of the density operator in the power of driving speed, we have identified the non-adiabatic correction to the quantum thermodynamic quantities in the strongly coupled regimes. With numerical tools, we have calculated non-adiabatic corrections for the resonant level model. We have verified our numerical results against analytical results for the zeroth and first order correction to the population in and out of equilibrium. We then have calculated the non-adiabatic entropy production out of equilibrium. We show that, at equilibrium, the friction, random force, and entropy production agree with each other. Out of equilibrium, friction, random force and entropy production are all correlation functions between different (generalized) random forces.  Our results agree with the one from the master equation in the limit of weak couplings. Future work must study the non-adiabatic entropy production and higher order corrections go beyond simple resonant level model.

\begin{acknowledgments}
We thank Jakob B\"atge for providing us data from HQME calculations. We also acknowledge the startup funding from Westlake University. 

\end{acknowledgments}

\section*{Data availability}
The data that support the findings of this study are available upon reasonable request from the authors.

\appendix
\section{Evaluation of  $\left<d\right|\d(\e-\hat{h})\left|d\right>$} \label{appA}
In this appendix, we draw certain identities and then evaluate $\left<d\right|\d(\e-\hat{h})\left|d\right>$. To start with, we have the following identity: 
\beq
\d(\e-\hat{h})(\e-\hat{h})=(\e-\hat{h})\d(\e-\hat{h})=0 .
\eeq
Taking ${\partial}_{\e_d}$ on both sides, we arrive at 
\beq
-{\partial}_{\e_d}\hat{h}\d(\e-\hat{h})+(\e-\hat{h}){\partial}_{\e_d}\d(\e-\hat{h})=-\d(\e-\hat{h}){\partial}_{\e_d}\hat{h}+{\partial}_{\e_d}\d(\e-\hat{h})(\e-\hat{h})=0 .
\eeq
Hence, 
\beq
{\partial}_{\e_d}\hat{h}=(\e-\hat{h})^{-1}{\partial}_{\e_d}\hat{h}\d(\e-\hat{h})=\d(\e-\hat{h}){\partial}_{\e_d}\hat{h}(\e-\hat{h})^{-1} .
\eeq
Similarly, we can show that 
\beq
{\partial}_{\e}\d(\e-\hat{h})=-(\e-\hat{h})^{-1}\d(\e-\hat{h})=-\d(\e-\hat{h})(\e-\hat{h})^{-1} .
\eeq
The above equations will be useful for higher order corrections. 

In the diagonal basis, we have 
\beq
\d(\e-\hat{h})=\sum_{k\a}\left|\psi_{k\a}\right>\d(\e-\e_{k\a})\left<\psi_{k\a}\right| ,
\eeq
such that 
\begin{eqnarray}
\left<d\right|\d(\e-\hat{h})\left|d\right>=\sum_{k\a}\left|\left<d|\psi_{k\a}\right>\right|^{2}\d(\e-\e_{k\a}) \nonumber \\
= \sum_{k\a} \frac{|V_{k\a}|^{2}\d(\e-\e_{k\a})}{(\e-\e_{d})^{2}+(\Gamma/2)^{2}}
=\frac{1}{2\pi} \sum_{\a}\frac{\Gamma_{\a}}{(\e-\e_{d})^{2}+(\Gamma/2)^{2}} =\frac{A(\e)}{2\pi}.  \label{ddd} 
\end{eqnarray}

\section{Deviation of Eq. $\ref{N2}$} \label{app:B}
At equilibrium, the one-body steady state density is 
\beq
\hat \sigma_{ss}=f(\hat{h})=\int f(\e)\d(\e-\hat{h})d\e \label{sig0},
\eeq
Use the identity in Appendix A, we can show that 
\beq
\hat \sigma^{(1)}=-\pi\hbar\dot{\e_d}\int d\e \d(\e-\hat{h}){\partial}_{\e_d}\hat{h}\d(\e-\hat{h}){\partial}_{\e}f(\e) \label{sigma1} ,
\eeq
Taking the derivative of  $\hat \sigma^{(1)}$ with respect to $\e_d$, we have
\begin{eqnarray}
{\partial}_{\e_d}\hat \sigma^{(1)}&=&-\pi\hbar\dot{\e_d}\int d\e {\partial}_{\e_d}\d(\e-\hat{h}){\partial}_{\e_d}\hat{h}\d(\e-\hat{h}){\partial}_{\e}f(\e) \nonumber \\
&&-\pi\hbar\dot{\e_d}\int d\e \d(\e-\hat{h}){\partial}_{\e_d}\hat{h}{\partial}_{\e_d}\d(\e-\hat{h}){\partial}_{\e}f(\e) \nonumber \\
&=&-\pi\hbar\dot{\e_d}\int d\e (\e-\hat{h})^{-1}{\partial}_{\e_d}\hat{h}\d(\e-\hat{h}){\partial}_{\e_d}\hat{h}\d(\e-\hat{h}){\partial}_{\e}f(\e) \nonumber \\
&&-\pi\hbar\dot{\e_d}\int d\e \d(\e-\hat{h}){\partial}_{\e_d}\hat{h}\d(\e-\hat{h}){\partial}_{\e_d}\hat{h}(\e-\hat{h})^{-1}{\partial}_{\e}f(\e)
\end{eqnarray}
We have used the identities shown in Appendix \ref{appA}. To evaluate $\hat \sigma^{(2)}$, we first calculate the following term, 
\begin{eqnarray}
&&\int_{0}^{\infty}e^{i\hat{h}t/\hbar}\int d\e (\e-\hat{h})^{-1}{\partial}_{\e_d}\hat{h}\d(\e-\hat{h}){\partial}_{\e_d}\hat{h}\d(\e-\hat{h}){\partial}_{\e}f(\e) e^{-i\hat{h}t/\hbar} \nonumber \\
&=&\sum_{mnp}\int_{0}^{\infty}dte^{i(\e_{m}-\e_{n})t/\hbar}\int d\e \left|m\right>(\e-\e_{m})^{-1}\left<m\right|{\partial}_{\e_d}\hat{h}\left|p\right>\d(\e-\e_{p})\left<p\right|{\partial}_{\e_d}\hat{h}\left|n\right>\d(\e-\e_{n}){\partial}_{\e}f(\e)\left<n\right| \nonumber \\
&=&\pi\hbar\sum_{mnp}\d(\e_{n}-\e_{m})\left|m\right>(\e_{n}-\e_{m})^{-1}\left<m\right|{\partial}_{\e_d}\hat{h}\left|p\right>\d(\e_{n}-\e_{p})\left<p\right|{\partial}_{\e_d}\hat{h}\left|n\right>{\partial}_{\e}f(\e_{n})\left<n\right| \nonumber \\
&=&\pi\hbar\sum_{mnp}\int d\e \left|m\right>(\e-\e_{m})^{-1}\d(\e-\e_{m})\left<m\right|{\partial}_{\e_d}\hat{h}\left|p\right>\d(\e-\e_{p})\left<p\right|{\partial}_{\e_d}\hat{h}\left|n\right>\d(\e-\e_{n}){\partial}_{\e}f(\e)\left<n\right| \nonumber \\
&=&-\pi\hbar \int d\e {\partial}_{\e}\d(\e-\hat{h}){\partial}_{\e_d}\hat{h}\d(\e-\hat{h}){\partial}_{\e_d}\hat{h}\d(\e-\hat{h}) {\partial}_{\e}f(\e) .
\end{eqnarray}
Similarly,
\begin{eqnarray}
&&\int_{0}^{\infty}e^{i\hat{h}t/\hbar}\int d\e \d(\e-\hat{h}){\partial}_{\e_d}\hat{h}\d(\e-\hat{h}){\partial}_{\e_d}\hat{h}(\e-\hat{h})^{-1}{\partial}_{\e}f(\e) e^{-i\hat{h}t/\hbar}  \nonumber  \\
&=&-\pi\hbar \int d\e \d(\e-\hat{h}){\partial}_{\e_d}\hat{h}\d(\e-\hat{h}){\partial}_{\e_d}\hat{h}{\partial}_{\e}\d(\e-\hat{h}) {\partial}_{\e}f(\e) .
\end{eqnarray}
Therefore, 
\beq
\begin{aligned}
\hat \sigma^{(2)}=&-\dot{\e_d}\int_{0}^{\infty}e^{i\hat{h}t/\hbar}{\partial}_{\e_d} \hat \sigma^{(1)}e^{-i\hat{h}t/\hbar} \\
=&-\pi^{2}\hbar^{2}\dot{\e_d}^{2}\int d\e {\partial}_{\e}\d(\e-\hat{h}){\partial}_{\e_d}\hat{h}\d(\e-\hat{h}){\partial}_{\e_d}\hat{h}\d(\e-\hat{h}) {\partial}_{\e}f(\e)\\
&-\pi^{2}\hbar^{2}\dot{\e_d}^{2}\int d\e \d(\e-\hat{h}){\partial}_{\e_d}\hat{h}\d(\e-\hat{h}){\partial}_{\e_d}\hat{h}{\partial}_{\e}\d(\e-\hat{h}) {\partial}_{\e}f(\e) .
\end{aligned}
\eeq
We then proceed to evaluate $N^{(2)}$:
\beq
\begin{aligned}
N^{(2)}=&\left<d\right|\sigma^{(2)}\left|d\right>=-\frac{\hbar^{2}}{4\pi}\dot{\e}_{d}^{2}\int d\e {\partial}_{\e}A(\e)A^{2}(\e){\partial}_{\e}{f}(\e)\\
=&-\frac{\hbar^{2}\dot{\e}_{d}^{2}}{12\pi}\int d\e {\partial}_{\e}A^{3}(\e){\partial}_{\e}{f}(\e)\\
=&-\frac{\hbar^{2}\dot{\e}_{d}^{2}}{12\pi}\left[A^{3}(\e){\partial}_{\e}{f}(\e) \right]_{0}^{\infty}+\frac{\hbar^{2}\dot{\e}_{d}^{2}}{12\pi}\int d\e A^{3}(\e){\partial}_{\e}^{2}{f}(\e)\\
=&\frac{\hbar^{2}\dot{\e}_{d}^{2}}{12\pi}\int d\e A^{3}(\e){\partial}_{\e}^{2}{f}(\e) ,
\end{aligned}
\eeq
which gives us the results shown in Eq. $\ref{N2}$. We have used the fact that $\partial_{\e_d}\hat{h} = | d \rangle \langle d |$.

\bibliography{e-ph_coupling_2}

\end{document}